\documentclass[fleqn,twoside]{article}
\usepackage{espcrc2}

\usepackage{graphicx}
\usepackage[figuresright]{rotating}

\newcommand{\AmS}{{\protect\the\textfont2
  A\kern-.1667em\lower.5ex\hbox{M}\kern-.125emS}}

\title{The X-ray Background and AGNs}

\author{G. Hasinger
        \address{ Max-Planck-Institut f\"ur extraterrestrische Physik \\
Postfach 1319, D--84541 Garching, Germany}}
       
\begin{document}

\begin{abstract}
Deep X--ray surveys have shown that the cosmic X--ray background (XRB) is largely due to the accretion onto supermassive black holes, integrated over the cosmic time. These surveys have resolved more than 80\% of the 0.1-10 keV X--ray background into discrete sources. Optical spectroscopic identifications show that the sources producing the bulk of the X--ray background are a mixture of unobscured (type-1) and obscured (type-2) AGNs, as predicted by the XRB population synthesis models. A class of highly luminous type-2 AGN, so called QSO-2s, has been detected in the deepest Chandra and XMM-Newton surveys. The new Chandra AGN redshift distribution peaks at much lower redshifts (z $\sim$ 0.7) than that based on ROSAT data, and the new X-ray luminosity function indicates that the space density of Seyfert galaxies peaks at significantly lower redshifts than that of QSOs. It is shown here, that the low redshift peak applies both to absorbed
and unabsorbed AGN and is also seen in the 0.5-2 keV band alone. 
Previous findings of a strong dependence of the fraction of type-2 AGN on luminosity are confirmed with better statistics here. Preliminary results from an 800 ksec XMM-Newton observation of the Lockman Hole are discussed.
\vspace{1pc}
\end{abstract}

\maketitle

\section{Introduction}

In recent years the extragalactic X--ray background in the 0.1-10 keV band
has almost completely been resolved into discrete sources with the deepest 
{\em ROSAT}, {\em Chandra} and {\em XMM-Newton} observations
\cite{has98,gia01,has01,ale03}. Optical identification programmes with 
Keck \cite{schm98,leh01,bar01,bar03}) and VLT \cite{szo03,fio03} find 
predominantly
unobscured AGN-1 at X-ray fluxes $S_X>10^{-14}$ erg cm$^{-2}$ s${-1}$, and a
mixture of unobscured and obscured AGN-2 at fluxes $10^{-14}>S_X>10^{-15.5}$
erg cm$^{-2}$ s$^{-1}$ with ever fainter and redder optical counterparts, while
at even lower X-ray fluxes a new population of star forming galaxies emerges
\cite{hor00,ros02}. At optical magnitudes R$>$24 all these 
surveys suffer, however, from spectroscopic incompleteness, so that 
photometric redshift techniques have to be applied here {\cite{zhe03}.

After having understood the basic contributions to the X-ray background, the
interest is now focusing on understanding the physical nature of these sources,
the cosmological evolution of their properties, and their role in models of
galaxy evolution. The X-ray observations have been roughly consistent with X-ray
background population synthesis assuming a mixture of absorbed and unabsorbed
AGN, folded with the corresponding luminosity function and cosmological
evolution, e.g.\cite{com95,fab98,gil01}. However, inputs to these models are 
still rather uncertain, like e.g. the cosmological evolution of low-luminosity 
AGN or the fraction of type-1 to type-2 AGN as a function of redshift and 
intrinsic luminosity, and a wide range of different assumptions has been
invoked for these parameters \cite{fra02,gan03}, see also \cite{gil03a}. 
Finally, the source statistics and optical incompleteness
are rather poor at high redhifts.

\begin{figure*}[t]
\includegraphics*[width=16cm]{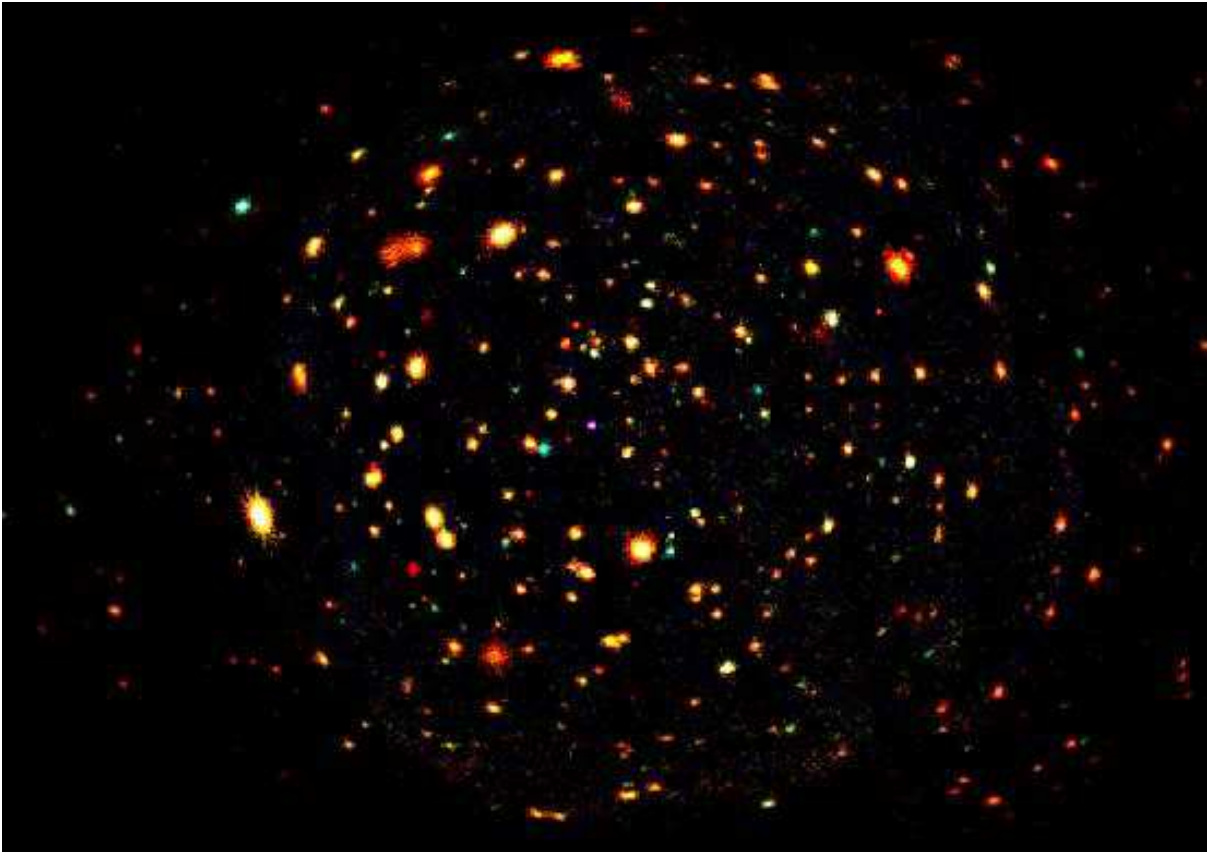}
  \vskip -1.0cm
  \caption{\small Color composite image of the $\sim$800 ksec XMM-Newton image 
  of the Lockman Hole. The image was obtained combining three energy bands: 
  0.5-2 keV, 2-4.5 keV, 4.5-10 keV (respectively red, green and blue), 
  following \cite{has01}. The image has a size of 
  $\sim$43$\times$30~arcmin$^2$.} 
\end{figure*}

The deep Chandra and XMM surveys, but also wider ASCA surveys have already
provided important new constraints. Several examples of the long-sought class of
high redshift, radio quiet type-2 QSO have been detected in deep fields 
\cite{nor02,ste02,bar03,szo03}. These allow for the first time to constrain the
fraction of type-2 AGN as a function of X--ray luminosity. At low luminosities a 
type-2 fraction of 75-80\% is found, consistent with local optically selected 
Seyfert galaxies, while at high luminosities the type-2 fraction is significantly 
smaller \cite{ued03,szo03}. The redshift distribution of Chandra deep survey 
sources peaks at z$\approx$0.7. This is related to the finding of a much slower 
cosmic evolution for Seyferts compared to QSOs \cite{ued03,has03,cow03,schm04,fio03}. 
And finally, significant spikes are found in the redshift distributions 
\cite{gil03,bar02}, indicating that AGN prefer to live in sheets of large-scale 
structure. 

In this review I 
show preliminary results on an ultradeep XMM-Newton observation in the Lockman
Hole and 
compare the optical identification work in the two deepest Chandra fields, 
the Chandra Deep Field South and the Hubble Deep Field North.

\section{XMM-Newton Ultradeep Survey of the Lockman Hole}

The Lockman Hole (LH) is the area with the absolutely lowest column density of 
interstellar absorption and was selected for the location of the deepest ROSAT 
surveys \cite{has98}, as well as the first XMM-Newton deep survey, observed in 
the performance verification phase of the satellite \cite{has01}. In AO2 a 200 
ksec observation of the LH was awarded to X. Barcons, but due to high background
conditions, unfortunately only about 50 ksec of good exposure time could be 
secured. In AO3, our XMM LH team, including X. Barcons and A. Fabian,
has been awarded an XMM exposure time of 665 ksec in the 
Lockman Hole. Due to a scheduling error,
the step width of the mini raster survey specified in order to reduce the 
imprint of the CCD chip gaps on the exposure map, has been set erroneously to 
15 arcminutes instead of 1 arcmin, so that a significant part of the exposure 
($>$ 300 ksec net) was spent in a wider raster pattern. While unfortunate
for the primary goal of our deep survey, this turned out to be very useful, 
because it provided a wider area coverage with an exposure time of almost 
100 ksec. Finally, the XMM-Newton observatory reimbursed the 300 ksec exposure
lost for the central pointing, so that the total net integrated XMM-Newton 
exposure in the Lockman Hole is about 800 ksec (see Figure 1).

\begin{figure*}[t]
  \vskip -1.0cm
  \includegraphics*[width=16.0cm]{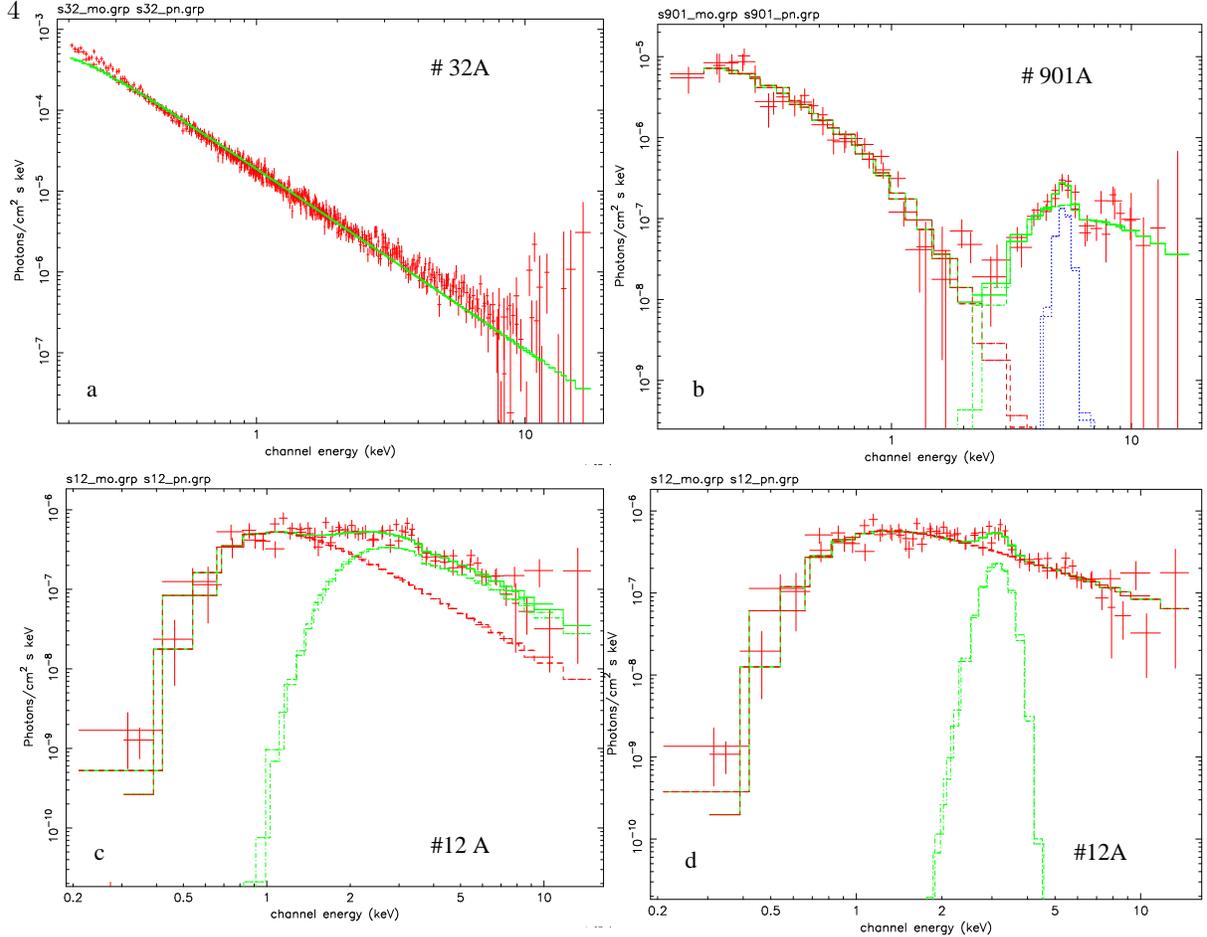}
  \vskip -1.0cm
  \caption{\small Selected X-ray spectra from the XMM-Newton observation of 
  the Lockman Hole. (a) ROSAT \#32A fit by a single power law without
  intrinsic absorption; (b) ROSAT \#901A fit by a soft thermal component
  plus a heavily absorbed power law with superposed, broad Fe emission 
  line; (c) ROSAT \#12A fit with a power law with two intrinsic absorber 
  thicknesses; (d) ROSAT \#12A fit with an intrinsically absorbed power law 
  and a broad Fe emission line.}
\end{figure*}

Figure 1 shows the colour composite XMM-Newton image of the Lockman Hole. 
North is up and East is to the left. This 
was constructed by combining images, smoothed with a Gaussian with 
$\sigma$=2'' in three bands (0.5-2 keV, 2-4.5 keV, 4.5-10 keV).  Blue sources 
are those undetected in the soft (0.5-2 keV) band, most likely due to 
intrinsic absorption from neutral hydrogen with column densities 
$N_H>10^{22}~cm^{-2}$. Very soft sources appear red.

\subsection{X-ray spectroscopy}

The large throughput and the unprecedented hard X-ray sensitivity of the 
telescopes aboard the XMM-Newton observatory allow for the first time to 
determine X-ray spectra of the faintest X-ray source population and constrain 
the evolution of their physical properties, in particular the X-ray absorption.
In the first deep survey taken with XMM-Newton in the Lockman Hole, 
it was possible to determine coarse X-ray spectra of about 50 X-ray sources
with more than 100 photons detected in the XMM observation, representing a 
largely complete, optically identified subsample \cite{mai02}. While type-1 
QSOs have the typical blue colours, type-2 sources follow much redder optical 
colour tracks expected for their host galaxy because the optical nucleus is 
obscured. X-ray spectrophotometry with XMM finds the expected strong 
correlation between X-ray absorption and optical obscuration \cite{mai02}. 
However, there are two high redshift type-1 QSOs which are optically unobscured
and X-ray absorbed, indicating possible differences in the dust properties
of these sources, e.g. \cite{gra97}. However, the existing samples are too 
small to make population studies. A 400 ksec XMM-Newton observation of
the Chandra Deep Field South (see below) is providing additional X-ray
spectroscopy samples \cite{str03}. In the new 800 ksec XMM observation 330 
sources have more than 200 counts and 104 have $>$ 1000 counts allowing to 
study emission and absorption properties of rather distant sources in 
great detail. The spectral diagnostics of high-redshift is in particular
helped by the extremely low neutral hydrogen column density towards the 
LH.

\begin{figure*}[t]
  \includegraphics*[width=16.0cm]{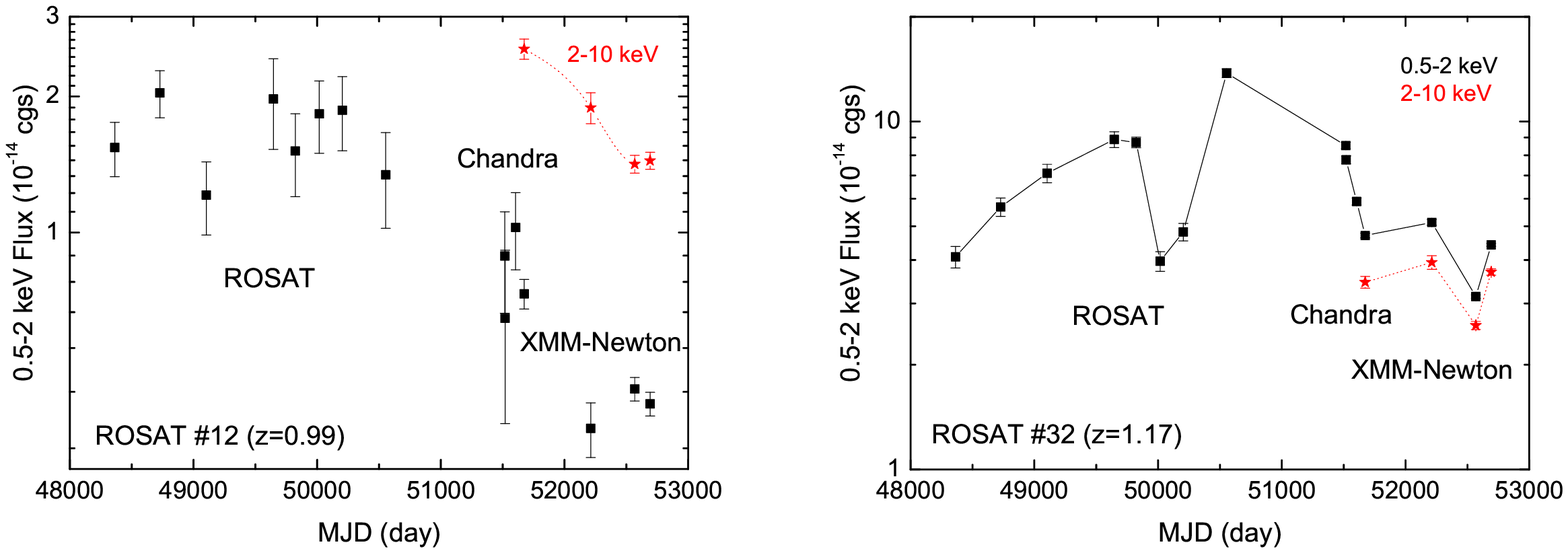}
  \vskip -1.0cm
  \caption{\small Long-term X-ray light curves of ROSAT \#12A (a) and 
   ROSAT \#32A (b) measured with several different satellites and 
   instruments. The first eight data points are taken from the
   ROSAT deep PSPC and ultradeep HRI observations of the Lockman 
   Hole \cite{has98}. The next three data points are from the Chandra HRC 
   observations of the Lockman Hole (PI: S. Murray). The last four 
   data points are from XMM-Newton (\cite{has01} and this work).
   Count rates have been converted to 0.5-2 keV fluxes assuming the spectral
   models shown in Figure 2. The short dashed line of data points 
   with star symbols gives the corresponding 2-10 keV fluxes from 
   XMM-Newton.} 
\end{figure*}

Figure 2 shows examples of unfolded X-ray spectra from three sources in the 
800 ksec observation of the LH, combining the data from the two 
MOS CCD and the pn-CCD camera of XMM-Newton. Panel 2a shows the spectrum
of ROSAT Deep Survey (RDS) source \#32A, a type-1 QSO at z=1.117, which can be 
fit to first order by a single power law model with photon index 
$2.18\pm0.02$ and galactic absorption. A detailed look at the residuals of
this fit, however, reveals excesses at soft energies (below 0.8 keV in
the QSO restframe), which could be associated with a soft excess. At 
high energies (above 8 keV in the QSO rest frame) there seems to be an
excess, which is likely associated to Compton reflection. No strong Fe 
line is apparent in this spectrum. Panel 2b shows the spectrum of ROSAT 
Ultradeep Survey (UDS) source \#901A, a relatively nearby Seyfert-2 galaxy 
at z=0.204. Its X-ray spectrum can be fit with a heavily obscured power law 
with $N_H=40\pm5 \cdot 10{22}$~cm$^{-2}$ and photon index $\sim 1.9$ and a superposed, 
broad Fe emission line, plus an additional unabsorbed soft thermal spectrum
with a temperature of $0.60\pm0.06$ keV, which, however does not show any 
emission lines. 

The spectrum of RDS source \#12A, a Seyfert-2 galaxy at z=0.99, is shown in 
panels 2c and 2d. It can be fit either with two heavily absorbed power law
spectra with a 'standard' photon index $\sim2$ and two absorber thicknesses 
of $N_H=2.3\cdot10^{22}$ and $2.6 \cdot 10^{23}$~cm$^{-2}$. This can be 
either interpreted (2c) as one AGN with partial covering absorption
with a covering factor of roughly 80\%, or even as a 
pair of AGN in the same galaxy, similar to the case recently discoverd in 
NGC 6240 \cite{kom03}. Alternatively it can be interpreted (2d) as a 
single, intrinsically absorbed power law with a rather flat photon index 
$\sim1.3$ and a superposed broad Fe emission line.

\subsection{Time variability}

One possibility to distinguish between different spectral model interpretations 
is to study the temporal variability of these very faint X-ray sources.
The Lockman Hole is ideally suited for this purpose, because it has been 
a deep Survey reference field for every imaging X-ray satellite following
ROSAT, i.e. ASCA, BeppoSAX, Chandra, and finally XMM-Newton, covering
a time base of about 13 years \cite{ish01,gio00,leh02,has01}. Figure 3
shows the light curves for two of the sources for which spectra are 
presented in Figure 2, the Seyfert-2 galaxy ROSAT \#12A and the 
QSO \#32A. The black points with error bars (almost invisible for \#32A)
give the 0.5-2 keV flux as a function of time, based on ROSAT PSPC, ROSAT 
HRI, Chandra HRC and XMM-Newton
EPIC data. It is surprising to see, that both the absorbed and the 
unabsorbed AGN show long-term time variations of up to a factor
of five. The red data points connected with dashed lines show 
the corresponding 2-10 keV flux (based only on XMM-Newton) for both sources, 
which varies significantly hand in hand with the soft X-ray flux. The second
XMM-Newton data point of \#12A, however, seems to deviate from this 
correspondence, which may indeed indicate that we are seeing two 
separate components varying independently. It would definitely be 
interesting to further monitor this source and to look for other,
similar examples of spectral and timing complexity.

The Lockman Hole contains a double-lobed high-redshift cluster of galaxies
(see Figure 1} in the North-East of the image) which was expected
to lie around a redshift of 1.26 \cite{hash02}. The new XMM-Newton data now 
allowed for
the first time to determine an X-ray redshift of z=1.13 from the 
Fe K$\alpha$ line of both lobes of the cluster. This is significantly 
different from the original redshift, which was based on a Keck 
NIR spectrum of a single central galaxy \cite{tho01}, but has later been 
confirmed by Keck spectroscopy of two fainter cluster galaxies. The 
XMM-Newton data can also constrain the chemical abundances of the 
intergalactic medium of the cluster, for the first time at such a
high redshift and rather low luminosity \cite{hash03}. 

With the new XMM catalogue in the Lockman Hole we already have been awarded
one night of spectroscopy with the powerful DEIMOS wide-field
spectrograph, in cooperation with P. Henry (IfA). The observations were 
taken in March 2003 and reduced by I. Lehmann (MPE). A total
of more than 100 spectra could be secured. However, due to the extremely
faint optical counterparts only 25 new redshift identifications were
obtained, raising the number of identified sources to about 135 in the LH.
Further spectroscopy at 4-10m class telescopes is necessary to complete
the optical spectroscopy in the Lockman Hole. However, for statistical 
studies a sample, which is rather complete for 0.5-2 keV fluxes 
down to $10^{-15}$~erg~cm$^{-2}$~s$^{-1}$ can be defined currently.

\section{The deepest Chandra Fields}

The Chandra X-ray Observatory has performed deep X-ray surveys in a number 
of fields with ever increasing exposure times \cite{mus00,hor00,gia01,ste02} 
and has  completed a 1 Msec exposure in the Chandra Deep Field South (CDF-S)  
\cite{ros02} and a 2 Msec exposure in the Hubble Deep Field North (HDF-N) 
\cite{ale03}. 

The CDF-S was also observed with XMM-Newton for a net exposure of ~400 ksec in 
July 2001 and January 2002 (PI: J. Bergeron, see \cite{has02}). The EPIC 
cameras have a larger field-of-view than ACIS, and a number of new diffuse 
sources are detected just outside the Chandra image. Similar to the LH, 
X-ray spectroscopy of a large number of sources will ultimately be very 
powerful with XMM-Newton \cite{str03}.

\begin{figure*}[t]
  \includegraphics*[width=16cm]{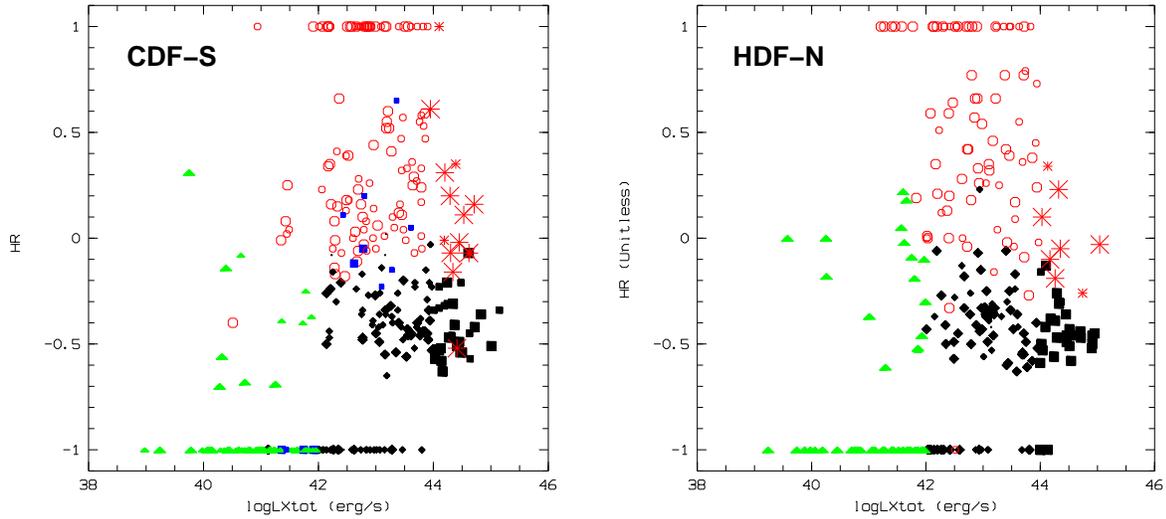}
  \vskip -1.0cm
  \caption{\small Hardness ratio versus rest frame luminosity in the total 
(0.5-10 keV) band. Objects are coloured according to their X-ray/optical 
classification: filled black diamonds correspond to low-luminosity type-1 AGN 
($log(L_X)<44$ erg s$^{-1}$), filled black squares to type-1 QSOs 
($log(L_X)>44$ erg s$^{-1}$), open red circles to low-luminosity type-2 AGN 
($log(L_X)<44$ erg s$^{-1}$), large red asterisks indicate type-2 QSOs
($log(L_X)>44$ erg s$^{-1}$), green triangles galaxies and blue squares 
extended X-ray sources. A consensus cosmology with 
$H_0$ = 70 km s$^{-1}$ Mpc$^{-1}$, $\Omega_m=0.3$ and $\Omega_\Lambda=0.7$ has 
been adopted. Luminosities are not corrected for possible intrinsic absorption.
 }
\end{figure*}

\subsection{VLT optical spectroscopy}

Optical spectroscopy in the CDF-S has been carried out in ~11 nights with the ESO Very Large Telescope (VLT) in the time frame April 2000 - December 2001, using deep optical imaging and low resolution multiobject spectroscopy with the FORS instruments with individual exposure times ranging from 1-5 hours. Some preliminary results including the VLT optical spectroscopy have already been presented \cite{nor02,ros02}. The complete optical spectroscopy is published in \cite{szo03}. 

Redshifts could be obtained so far for 169 of the 346 sources in the CDF-S, of which 144 are very reliable (high quality spectra with 2 or more spectral features), while the remaining optical spectra contain only a single emission line, or are of lower S/N. For objects fainter than R=24 reliable redshifts can be obtained if the spectra contain strong emission lines. For the remaining optically 
faint objects, and those not covered by the multiobject spectroscopy, we have 
to resort to photometric redshift techniques \cite{wol01,mai03,zhe03}. 
For a subsample at off-axis angles smaller than 
10 arcmin we obtain a spectroscopic completeness of about 50\%. Including 
photometric redshifts this completeness increases to $\approx$ 96\%. 

\begin{figure*}[t]
  \includegraphics*[width=16cm]{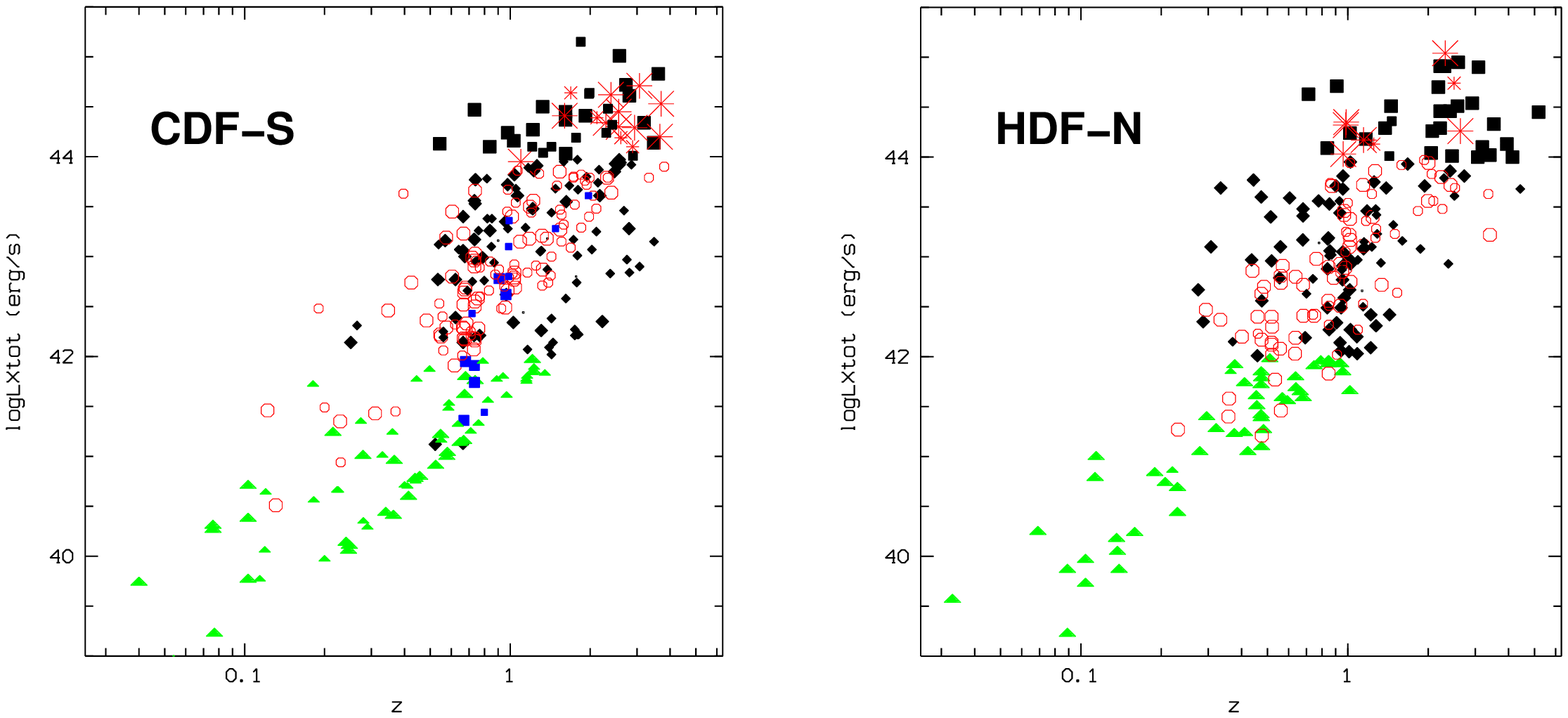}
  \vskip -1.0cm
  \caption{\small X-ray luminosity versus redshift magnitude for the CDF-S-sources (left) and the HDF-N sources (right). Symbols are the same as in Figure 4.}
\end{figure*}

\subsection{X-ray/optical classification}

Following \cite{ros02,has02} we show in Figure 4 {left} the hardness ratio as a function of the luminosity in the 0.5-10 keV band for 169 sources for which we have optical spectra and rather secure classification in the CDF-S \cite{szo03}
(large symbols) and about 170 objects with photometric redshifts \cite{zhe03}
(small symbols).  The X-ray luminosities are not corrected for internal absorption and are computed in a 'consensus cosmology' with $H_0$ = 70 km s$^{-1}$ Mpc$^{-1}$, $\Omega_m=0.3$ and $\Omega_\Lambda=0.7$. Different source types are clearly segregated in this plane. Type-1 AGNs (black diamonds) have luminosities typically above $10^{42}$ erg s$^{-1}$, with hardness ratios in a narrow range around HR$\approx$-0.5. Type-2 AGN are skewed towards significantly higher hardness ratios (HR$>$0), with (absorbed) luminosities in the range $10^{41-44}$ erg s$^{-1}$. Direct spectral fits of the XMM-Newton and Chandra spectra clearly indicate that these harder spectra are due to neutral gas absorption and not due to a flatter intrinsic slope (see \cite{mai02,bau03}). Therefore the unabsorbed, intrinsic luminosities of type-2 AGN would fall in the same range as those of type-1's. 

In Figure 4 we also indicate the type-2 QSOs (asterisks), the first one of which was discovered in the CDF-S \cite{nor02}. In the meantime, more examples have been found in the CDF-S and elsewhere \cite{ste02}. It is interesting to note that no high-luminosity, very hard sources exist in this diagram. This is a selection effect of the pencil beam surveys: due to the small solid angle, the rare high luminosity sources are only sampled at high redshifts, where the absorption cut-off of type-2 AGN is redshifted to softer X-ray energies. Indeed, the type-2 QSOs in this sample are the objects at $L_X>10^{44}$ erg s$^{-1}$ and HR$>$-0.2. The type-1 QSO in this region of the diagram is a BAL QSO with significant intrinsic absorption.

About 10\% of the objects have optical spectra of normal galaxies (marked with triangles), luminosities below $10^{42}$ erg s$^{-1}$ and very soft X-ray spectra (several with HR=-1), as expected in the case of starbursts or thermal halos. 
In these galaxies the X-ray emission is likely due to a mixture of 
hot, thermal gas and a population of low mass X-ray binaries \cite{bar01},
plus possibly some low level AGN activity \cite{nor03}.  
Therefore the deep Chandra and XMM-Newton surveys detect for the first time the population of normal starburst galaxies out to intermediate redshifts \cite{mus00,gia01,leh02}. These galaxies might become an important means to study the star formation history in the universe completely independently from optical/UV, sub-mm or radio observations \cite{nor03}.

Figure 4 (right) shows the same diagram for the spectroscopic and photometric
identifications in the 1Msec catalogue of the HDF-N \cite{bar02,bar03}. While 
the authors give only purely optical classification information for the X-ray counterparts (basically ''galaxy'' or ''broad line object''), we have applied the above X-ray/optical classification scheme also to their catalogue. The corresponding diagram shows basically the same features: the X-rays show that type-1 (mainly broad-line) AGN cluster around HR$\approx$-0.5 and break the degeneracy between type-2 Seyferts and normal galaxies.

The CDF-S identification catalogue \cite{szo03} includes 8 type-2 QSOs at 
redshifts above 2 (see figure 3 left), which all are characterised by strong 
and narrow UV emission lines (Lyman-$\alpha$, CIV etc.) with almost absent 
continuum. Additionally, there are three objects with photometric redshifts,
luminosities and hardness ratios consistent with type-2 QSOs., The 
corresponding HDF-N catalogue \cite{bar02} lists only two possible type-2 
QSOs at such high redshifts and one additional photometric candidate 
\cite{bar03}. A closer look at Figure 5 and the optical magnitudes of the CDF-S type-2 QSO shows that these objects are predominantly detected at optical magnitudes R$>$ 24. One reason for the relative absence of this population in the HDF-N could be that a smaller number of identifications at R$>24$ exist in this survey compared to the CDF-S. However, a true field to field variation (cosmic variance) or statistical fluctuation cannot be ruled out.

\begin{figure*}[t]
  \includegraphics*[width=16.0cm]{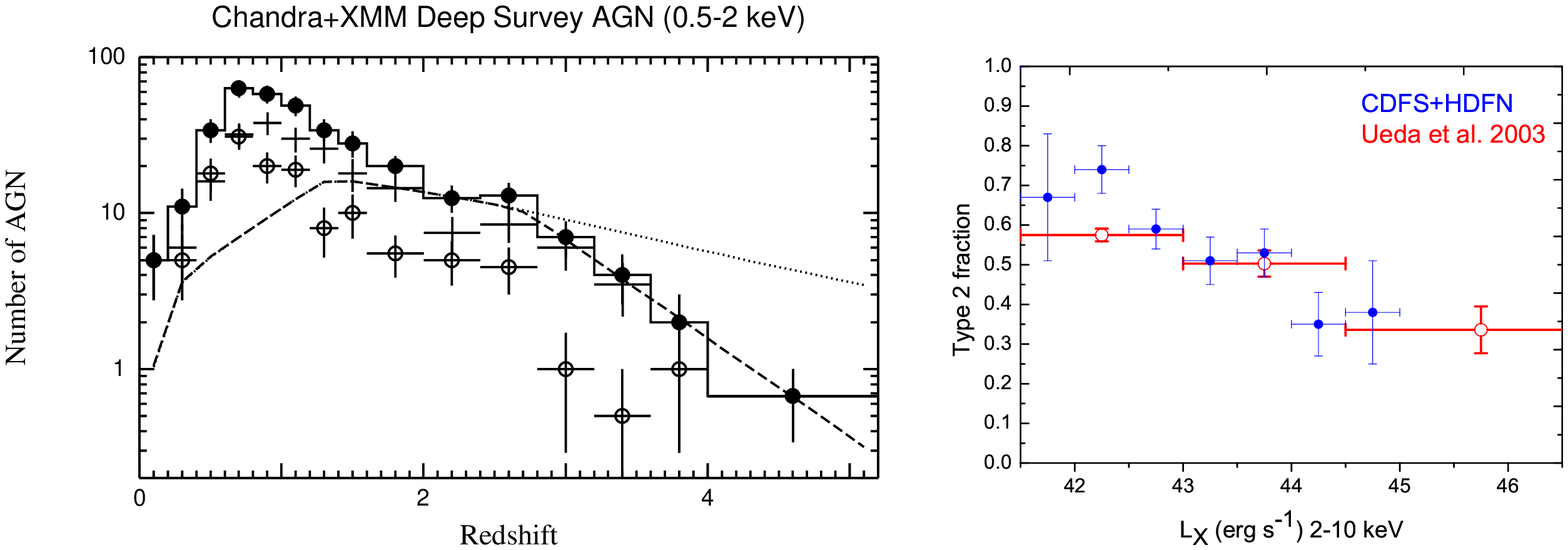}
  \vskip -1.0cm
  \caption{\small Left: Redshift distribution of 403 AGN selected in the 0.5-2 keV
band from the Chandra CDF-S and HDF-N and the XMM-Newton Lockman Hole survey
samples (solid circles and histogram), compared to model predictions from
population synthesis models \cite{gil01}. The dashed line shows the prediction
for a model, where the comoving space density of high-redshift QSO follows the
decline above z=2.7 observed in optical samples \cite{ssg95,fan01}. The
dotted line shows a prediction with a constant space density for $z>1.5$. The
two model curves have been normalized to fit the observed distribution
in the redshift range z=2-2.4. Simple crosses and open circles give the
redshift distribution separately for 252 AGN-1 and 151 AGN-2, respectively.
Right: fraction of type-2 AGN among all AGN for the CDF-S and HDF-N AGN
selected in the 2-10 keV band (solid blue symbols) and those given in
\cite{ued03} (open red symbols).}
\end{figure*}

\section{Redshift and luminosity distributions for different AGN types}

The current spectroscopic/photometric completeness of the CDF-S and HDF-N identifications allows to compare the observed redshift distribution with predictions from X--ray background population synthesis models, which, due to the saturation of the QSO evolution predict a maximum at redshifts around z=1.5. Figure 6 shows two predictions of the redshift distribution from the Gilli et al. model \cite{gil01} for a flux limit of $2.3 \times 10^{-16}$ erg cm$^{-2}$ s$^{-1}$ in the 0.5-2 keV band with different assumptions for the high-redshift evolution of the QSO space density. The model curves have been normalized to fit the observed 
distribution in the redshift range 2-2.4. 

The actually observed redshift distribution of AGN selected from the HDF-N and 
CDF-S Chandra deep survey samples at off-axis angles below 10 arcmin and in the
0.5-2 keV band is shown in Fig. 5 (left) as histogram and data points (solid
circles). In the 
redshift range below 1.5 it is radically different from the prediction, with a 
peak at redshift z$\approx$0.7. This low redshift peak is dominated by 
Seyfert galaxies with X-ray luminosities in the range $L_X = 10^{42-44}$ 
erg s$^{-1}$. 
This figure also shows, that the redshift distributions separated into type-1 
AGN (crosses) and type-2 AGN (crosses with open circles) both peak
significantly below z=1, which is not consistent with some recent 
population models, where the low redshift peak was assumed to be mainly 
in the type-2 AGN population \cite{fra02,gan03}.
  
Since the peak in the observed redshift distribution is expected at the 
redshift, where the strong positive evolution of AGN terminates, we can 
conclude that the evolution of Seyfert galaxies is significantly different 
from that of QSOs, with their evolution saturating around a redshift of 0.7, 
compared to the much earlier evolution of QSOs which saturates at 
z$\approx$1.5 \cite{miy00}. The statistics of the two samples is now sufficient to rule out 
the constant space density model at redshifts above 3, clearly indicating a 
decline of the X-ray selected QSO population at high redshift consistent with 
the optical findings.

Figure 6 (right) shows the ratio of type-2 AGN to all AGN selected 
in the 2-10 keV band for both the CDF-S and HDF-N (spectroscopic as
well as photometric redshifts) as a function of the 2-10 keV luminosity. As 
already indicated by Ueda et al. \cite{ued03} and confirmed by
Szokoly et al. \cite{szo03} there is a strong dependence of the 
fraction of type-2 AGN, i.e. the ratio of type-2 over all AGN
as a function of luminosity. While the type-2 fraction at low luminosities
is consistent with the local ratio of optically selected Seyfert 
galaxies \cite{huc92} of about 4:5, at high luminosities the 
ratio is 1:3. This indicates a breakdown of the 'strong unification'
model, where the covering factor is independent of luminosity and 
redshift. A possible explanation of this trend may be, that high 
luminosity objects are able to 'clean out' their environment
by ionizing the circumnuclear matter and/or producing strong
outflows, while low luminosity objects are largely surrounded 
by the circumnuclear starburst region they are feeding from
(see e.g. \cite{fab99}).

These, still preliminary, new results paint a dramatically different evolutionary picture for low-luminosity AGN compared to the high-luminosity QSOs. While the rare, high-luminosity objects can form and feed very efficiently rather early in the universe, the bulk of the AGN has to wait much longer to
grow. This could indicate two modes of accretion and black hole growth with different accretion efficiency, as e.g. proposed in \cite{dus02}.  
The late evolution of the low-luminosity Seyfert population is very similar to 
that which is required to fit the Mid-infrared source counts and background (see e.g. \cite{fra02}), however, contrary to what has been assumed by Franceschini et al., this evolution applies to all low-luminosity AGN (type-1 and type-2). 

These results, however, have still to be taken with a grain of salt. First, the spectroscopic incompleteness in the Chandra samples is still substantial and before a formal publication of all spectroscopic and photometric redshifts is available, it is too early to draw final conclusions.

\section*{Acknowledgments} 

I thank my colleagues in the the Chandra Deep Field South team and the Lockman 
Hole team, but in particular X. Barcons, J. Bergeron, H. B\"ohringer, 
A. Fabian, R. Giacconi, R. Gilli, P. Henry, I. Lehmann, V. Mainieri, C. Norman, 
P. Rosati, M. Schmidt, G. Szokoly, W. Zheng, for the good cooperation and the 
permission to use some data in advance of publication.

\end{document}